\documentclass[pdflatex,sn-mathphys-num]{sn-jnl}

\usepackage{graphicx}%
\usepackage{multirow}%
\usepackage{amsmath,amssymb,amsfonts}%
\usepackage{amsthm}%
\usepackage{mathrsfs}%
\usepackage[title]{appendix}%
\usepackage{xcolor}%
\usepackage{textcomp}%
\usepackage{manyfoot}%
\usepackage{booktabs}%
\usepackage{algorithm}%
\usepackage{algorithmicx}%
\usepackage{algpseudocode}%
\usepackage{listings}%

\begin{document}

\title[Article Title]{The simulation on neutron background reduction for InDEx at JUSL}

\author[1,2]{\fnm{Susmita} \sur{Das}}\email{susmita.das@saha.ac.in}

\author[1,2]{\fnm{Maitreyee} \sur{Nandy}}\email{maitreyee.nandy@saha.ac.in}

\author*[1,2]{\fnm{Mala} \sur{Das}}\email{mala.das@saha.ac.in}

\affil*[1]{\orgname{Saha Institute of Nuclear Physics}, \orgaddress{\street{1/AF, Bidhannagar}, \city{Kolkata} \postcode{700064}, \state{West Bengal}, \country{India}}}

\affil[2]{\orgname{Homi Bhabha National Institute}, \orgaddress{\street{Training School Complex, Anushakti Nagar}, \city{Mumbai} \postcode{400094}, \state{Maharashtra}, \country{India}}}

\abstract{Dark matter experiments are rare event search experiments that require zero background environment over very long exposures. To achieve this condition, a detailed simulation of detector geometry and experimental setup is required before the experiment is executed. Simulation plays a significant role in detector design and also provides a cost-effective and risk-free approach for predicting outcomes before real world experimentation. The present simulation work is focused on neutron background reduction for a dark matter direct detection experiment in India, the \underline{In}dian \underline{D}ark matter search \underline{Ex}periment (InDEx). The FLUKA and FLAIR simulation tools have been used throughout the simulation process. The experimental and simulation results available in the literature are being reproduced using FLUKA for validation purposes. The calibration and InDEx experiment are simulated, and the results are compared against the experimental results. For neutron background reduction in future experiments, the use of high density polyethylene (HDPE) is suggested and a shielding design using HDPE is presented. The results show that shielding reduces detector event rates by two orders of magnitude compared to the prior InDEx experiment without shielding.}

\keywords{Simulation, FLUKA, Neutron shielding, InDEx}

\maketitle

\section{Introduction}\label{sec1}

For rare event detection, background identification and suppression are the main fundamental problems. In some cases, the background events are difficult to differentiate from the expected signal as they mimic the signal event. This happens because the detector responds to nuclear and electron recoils produced by the incident particles. For example, in dark matter detection, the neutrons and gamma-rays induced recoils are similar to WIMP-induced recoils. Thus, in order to detect the actual events aimed to be studied and to achieve sufficient experimental sensitivity, measurements need to be carried out in a reduced or identified background environment. This necessitates suppression of the background events by passive or active shielding. In some cases, rejection of a certain background is also possible by choosing suitable detector technologies. In the field of radiation physics, the Superheated Emulsion Detector (SED) has been utilized for an extended period of time \cite{ROY2001271}. It has several applications, including neutron detection, neutron dosimetry, gamma-ray detection, proton and heavy ion detection, as well as neutron spectrometry. The major advantage of SED is its threshold-dependent background rejection capability, where the threshold of the detector varies with the operating temperature and pressure. Below a certain threshold, some backgrounds (e.g., gamma/electrons) can be rejected and that makes this type of detector suitable for rare event detection like dark matter searches. Dark matter experiments, e.g., SIMPLE, COUPP, MOSCAB, PICASSO and PICO \cite{PhysRevLett.108.201302, fustin2024darkmatterlimitscoupp, Antonicci2017, BEHNKE201785, PhysRevD.100.022001} have already used superheated liquid in droplet detector and as the bubble chamber. These superheated liquid detectors are aimed to probe dark matter candidates, WIMPs, in the mass region of sub-GeV to GeV. \underline{In}dian \underline{D}ark matter search \underline{Ex}periment (InDEx) at \underline{J}aduguda \underline{U}nderground \underline{S}cience \underline{L}aboratory (JUSL) uses superheated liquid detectors in the form of a droplet detector \cite{vpb7-kdrz}. 

The origin of the backgrounds (intrinsic or external) depends on the detector materials, the experimental setup, the subterranean depth and the environment of the laboratory. At the surface laboratory, the major external background is the cosmic rays, whereas in the underground, the cosmic flux is reduced significantly following the exponential reduction in muon flux with the vertical depth \cite{PhysRevD.73.053004,PhysRevD.110.063006}. For example, the muon flux at JUSL is reduced by four orders of magnitude compared to the surface \cite{SHARAN2021165083}. The rock overburden at the depth of the laboratory acts as a natural shielding for the cosmic ray background. This is the motivation behind the establishment of deep underground facilities, such as SNOLAB (Canada), LSM (France), SURF (U.S.), LNGS (Italy), SOUDAN (U.S.), Y2L (China) and JUSL (India) etc. In underground environments, the natural radioactivity within the Earth's crust contributes to the radiogenic background. The radiation backgrounds present there consist primarily of neutrons, gamma-rays, radon, alpha particles and electrons. The high energy neutrons are generated by the spallation reaction of highly penetrating cosmic muons with underground rock or by a hadronic shower. On the other hand, the radioisotopes (U, Th, K, Rn, etc.) present in rock produce alpha, beta, and gamma-rays through the radioactive decay process. These radioisotopes, being primordial and long-lived, act as an irreducible source of background in detector materials as well as in the environment. The neutrons (low energy) are produced by the radioactive decay process and by the ($\alpha$, n) reaction.

The direct detection experiment, InDEx located at JUSL in India, uses superheated liquid of C\(_2\)H\(_2\)F\(_4\) as the detector active target. The gamma-sensitivity of that superheated liquid starts above 38\(^{\circ}\)C \cite{SAHOO201944, SAHOO2021165450}. The experimental runs below this gamma-sensitive region have already been carried out \cite{KUMAR2026171101, vpb7-kdrz}. In those small-scale runs of the InDEx, no shielding against the background was incorporated. It is shown that under zero background, InDEx can achieve significantly improved detector sensitivity, approximately three orders of magnitude lower than the current limit \cite{vpb7-kdrz,PhysRevD.101.103005}. Therefore, at this stage of experimental progress of InDEx, it is necessary to design a proper background shielding. This paper will represent the first step taken in this direction.

The underground facility JUSL is situated at 555 m (1604 m water equivalent) beneath the surface in Jharkhand, India. An earlier background measurement at JUSL \cite{GHOSH2022102700} confirms the presence of gamma-rays and neutrons as major backgrounds. The present work is focused on the passive shielding design for the neutron background at JUSL. Since the experiments are currently planned to be conducted below the gamma-sensitive threshold, the primary concern is the neutron background. The neutron contribution comes from both radiogenic and cosmogenic sources. The radiogenic neutrons are mainly produced through ($\alpha$, n) reaction and through the spontaneous fission of Uranium-238 in the surrounding rock. The cosmogenic neutrons are produced by several processes, like muon capture, muon spallation reaction, and muon-induced hadronic and electromagnetic cascades. The measured radiogenic neutron flux at JUSL is \((1.61\pm0.03) \times 10^{-4}\) cm\(^{-2}\) s\(^{-1}\) where \((9.93\pm0.22\pm0.10) \times 10^{-5}\) cm\(^{-2}\) s\(^{-1}\) is the fast neutron flux and \((6.15\pm0.18) \times 10^{-5}\) cm\(^{-2}\) s\(^{-1}\) is the thermal neutron flux \cite{GHOSH2022102700}. The simulation of the cosmogenic neutrons results in a very low contribution of neutron flux \((8.458\pm0.826_{stat}\pm0.003_{sys}) \times 10^{-8}\) cm\(^{-2}\) s\(^{-1}\) with energies up to a few GeVs \cite{GHOSH2022102700}. Designing a shielding for the experiment requires simulations for material selection and thickness optimization. Currently, the shielding will be designed for InDEx operation at 1.95 keV of Seitz threshold of the detector. This threshold is chosen because, in run 2 of InDEx, the experiment was carried out at 35\(^{\circ}\)C, which corresponds to 1.95 keV thresholds and no shielding was used \cite{vpb7-kdrz}. In the present work, to improve the detector sensitivity over the run 2 results, the shielding design is simulated. In the future, this neutron shielding will be incorporated with other background shielding, such as for gamma-rays, as the detector becomes sensitive to gamma-rays at higher temperatures ($>$ 38\(^{\circ}\)C) or lower thresholds. The minimum neutron energy that can be detected at this 1.95 keV threshold is 7 keV and 11 keV for carbon and fluorine nuclei, respectively, in the  C\(_2\)H\(_2\)F\(_4\) target \cite{vpb7-kdrz}. The simulation is performed with the FLUKA \cite{refId0,BOHLEN2014211} particle interaction and transport code and FLAIR \cite{refId02} simulation toolkit. The first section of the manuscript discusses the use of the SED in dark matter searches and the radioactive backgrounds found at JUSL. In the second section, an overview of the methodology is discussed, along with the validation of FLUKA results. Section three discusses a detailed simulation study of the calibration and InDEx run 2 experimental setup. The manuscript is concluded with the findings and discussion, followed by a conclusion. 

\section{Methods}\label{sec2}

In this work, the shielding design for neutrons has been simulated using FLUKA. FLUKA is a Monte Carlo simulation code used in various fields, including particle physics, nuclear physics, medical physics, radiation protection etc. The present study is centred on neutron transmission through different materials. To specify the neutron source, the SOURCE routine has been employed. The transmitted neutron fluence has been calculated using the surface crossing fluence calculator USRBDX. The particular geometry and source that are considered for different simulations are discussed in detail in the respective sections.

\begin{figure}
\centering
\includegraphics[width=0.9\textwidth]{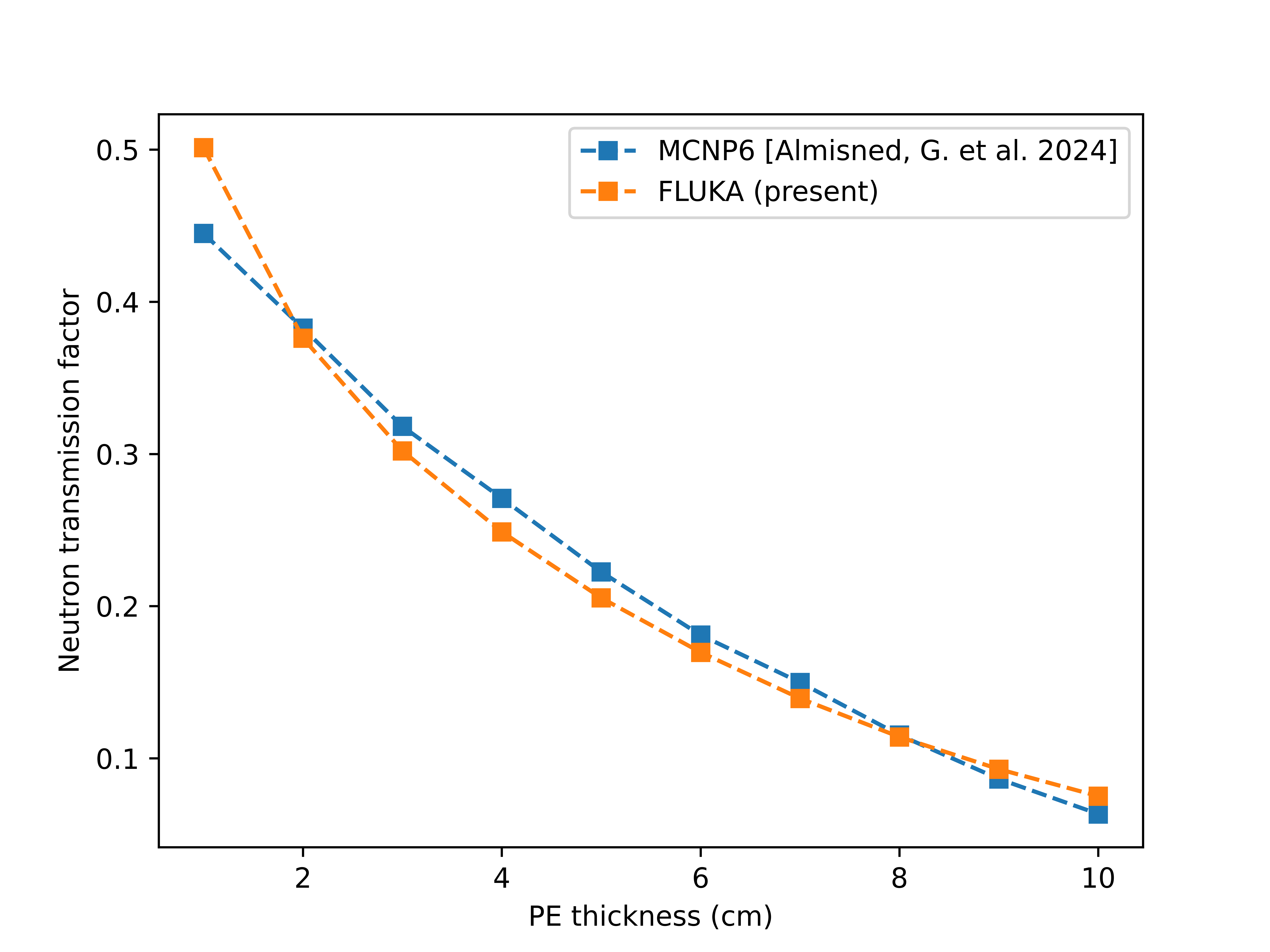}
\caption{Neutron transmission simulated with FLUKA and MCNP6 results [Almisned, G. et al. 2024]. \cite{ALMISNED2024111585}}\label{fig1}
\end{figure}

\begin{figure}
\centering
\includegraphics[width=0.9\textwidth]{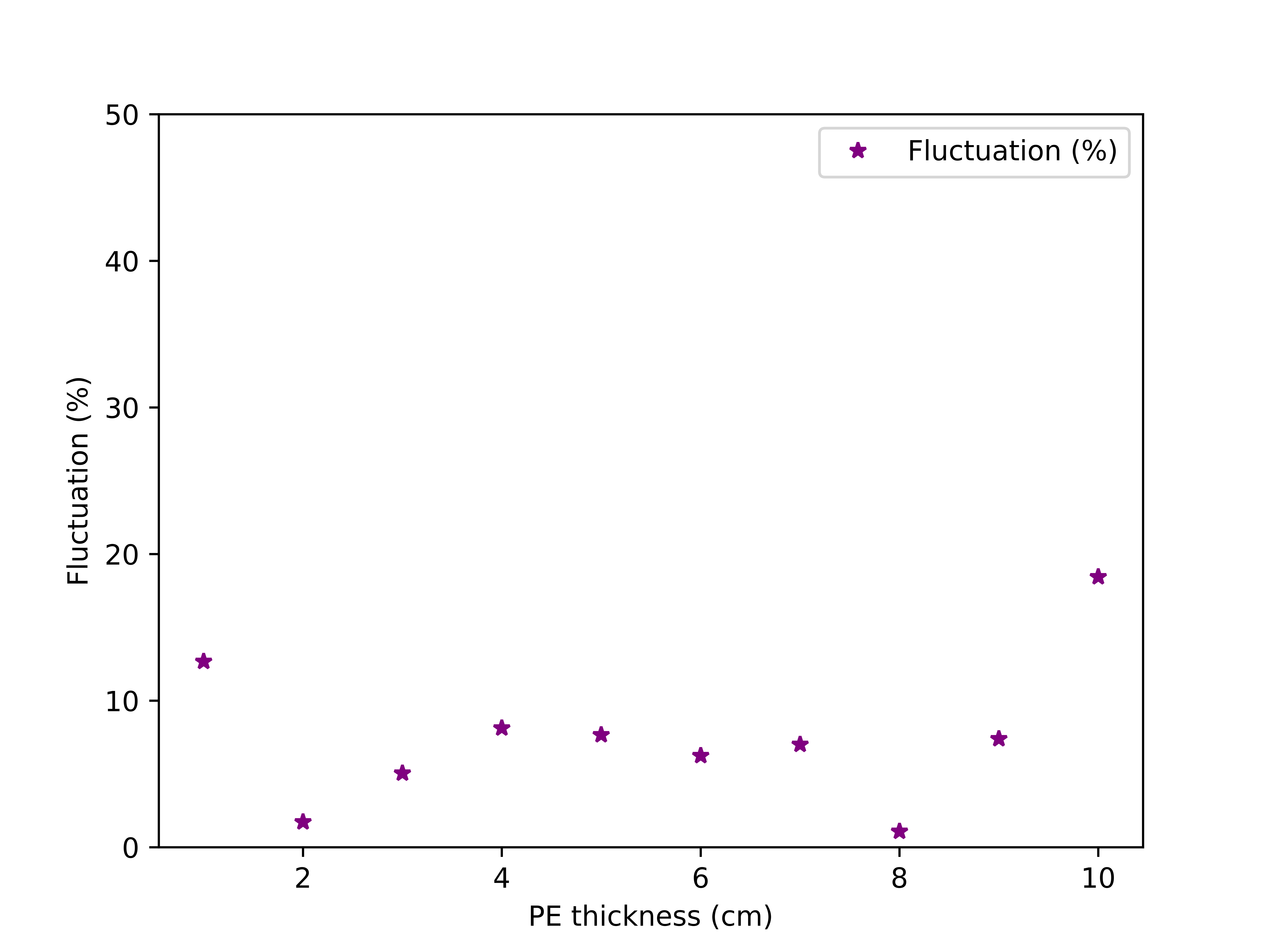}
\caption{Percentage variations between MCNP6 and FLUKA results at different PE thicknesses.}\label{fig2}
\end{figure}

\begin{figure}
\centering
\includegraphics[width=0.9\textwidth]{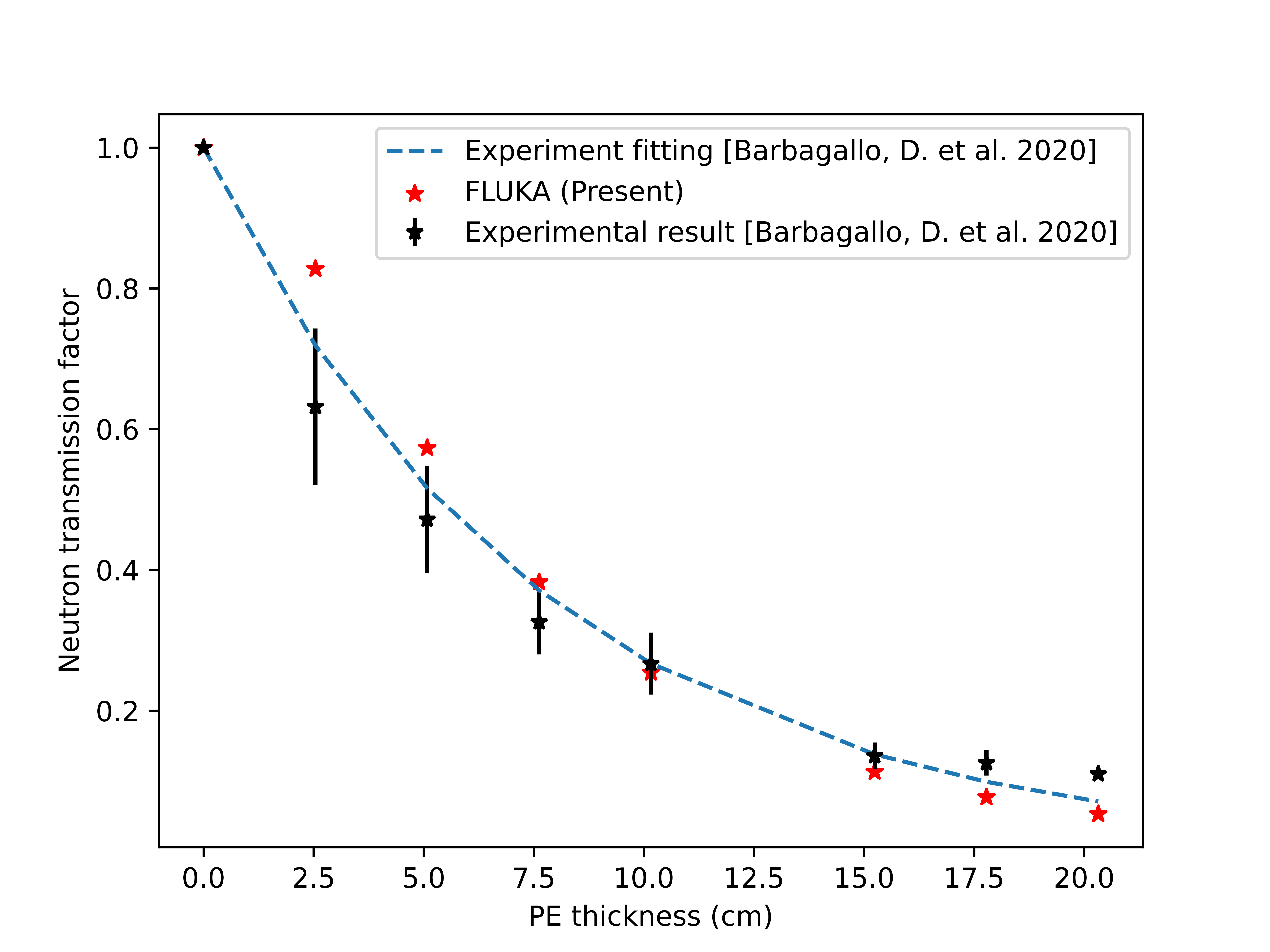}
\caption{Comparison of neutron transmission simulated using FLUKA with Experimental results [Barbagallo, D. et al. 2020] \cite{10.1063/10.0002041}.}\label{fig3}
\end{figure}

In order to validate the results obtained through simulations with FLUKA, it has been made to reproduce two results from existing literature. In the first case, the neutron transmission factor has been calculated for different thicknesses (0-10 cm) of polyethylene (PE), considering a 4.5 MeV monoenergetic isotropic neutron source. The geometry and source specification are followed according to the geometry mentioned in Ref. \cite{ALMISNED2024111585}, where the MCNP6 simulation code is used. The outcomes from the FLUKA (present) and MCNP6 simulations \cite{ALMISNED2024111585} for varying polyethylene thicknesses are illustrated in Fig. \ref{fig1}. The percentage variations between MCNP6 and FLUKA results at different PE thickness is shown in Fig. \ref{fig2}. The second study is about the simulation of an experiment with a bubble detector as mentioned in Ref \cite{10.1063/10.0002041}. Here, the neutron source is an isotropic polyenergetic (200 keV- 10 MeV) \(^{241}\)AmBe source. The SOURCE routine is used to define the \(^{241}\)AmBe spectra in FLUKA. The neutron transmission factor for various polyethylene thicknesses (0-20 cm) is calculated. A comparison plot between the present FLUKA results and the experimental data \cite{10.1063/10.0002041} is illustrated in Fig. \ref{fig3}.

The results in the first case show that the findings from the present work (FLUKA) and MCNP6 \cite{ALMISNED2024111585} appear to be consistent, except for the 1 cm and 10 cm polyethylene thicknesses. At these thicknesses, the discrepancies between the FLUKA values and the MCNP6 values are 13\% and 18\%, respectively. For all other results, the variation is within 10\%. In the second case, the simulated outcomes align with the experimental results for the majority of thicknesses. The discrepancies with the experimental data are within a 5\% error margin for six thicknesses, while at a thickness of 2.54 cm, the error reaches 11\%, and at 20.32 cm, the largest error is 24\%. Overall, the present simulation reproduces the trends of the available results from the literature.

\begin{figure}
\centering
\includegraphics[width=0.9\textwidth]{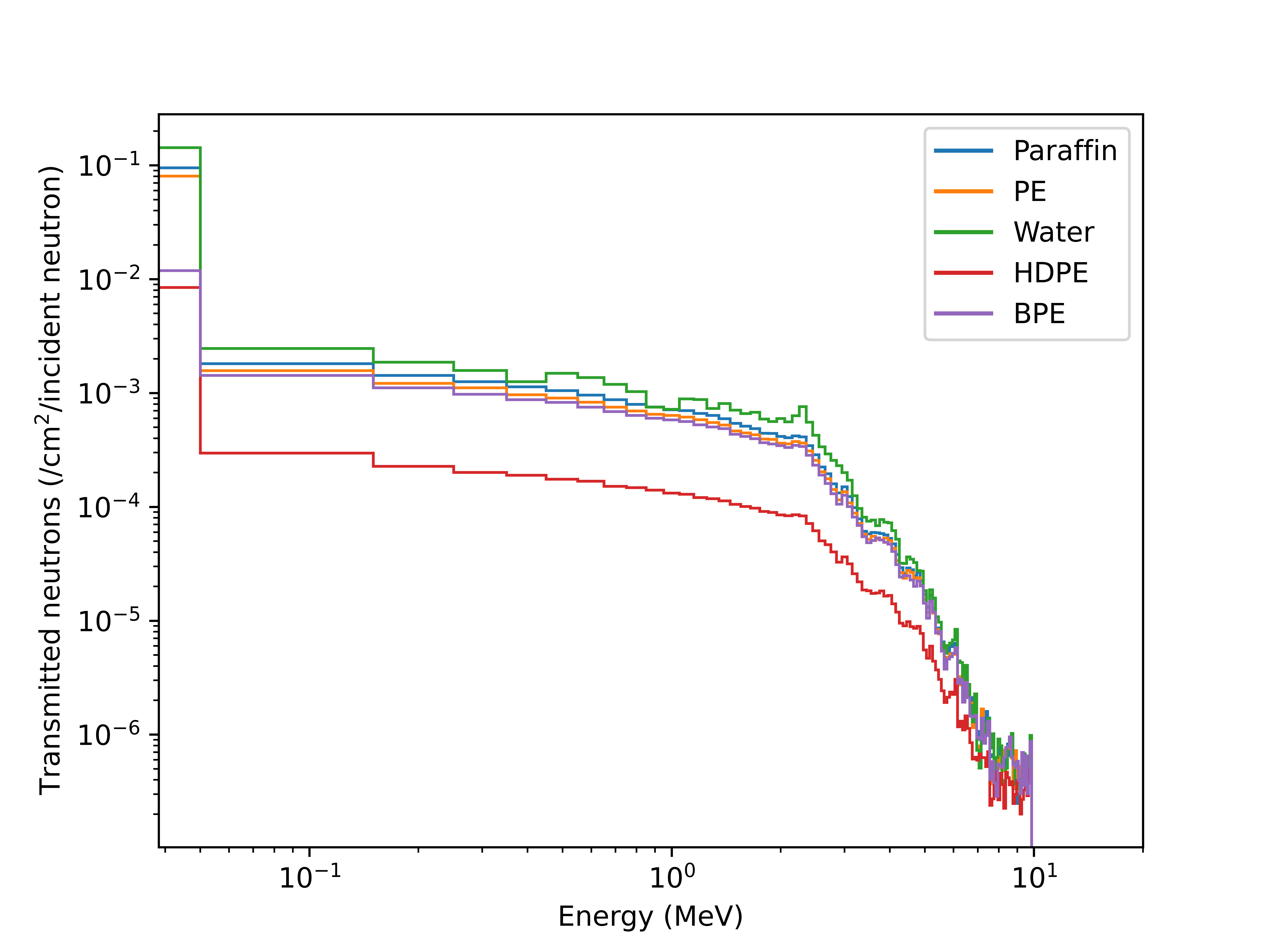}
\caption{Transmitted neutron spectra through 10 cm} thick, for JUSL radiogenic neutrons.\label{fig4}
\end{figure}

The most effective materials for neutron shielding are typically those rich in hydrogen, including water, paraffin, polyethylene, or concrete. The hydrogen present in materials significantly slows down neutrons. The fundamental principle of neutron moderation is based on energy loss via scattering with target nuclei. Kinematically, the energy transfer per collision is maximised when the target nucleus has a mass comparable to that of neutron. Hydrogen, with a mass nearly equal to that of a neutron, enables maximum energy transfer. Furthermore, its large scattering cross section facilitates efficient neutron moderation over short distances. Boron-containing compounds are also used to improve neutron absorption. InDEx detectors are insensitive to thermal neutrons \cite{article}. Here, different materials, such as water, PE, high density polyethylene (HDPE), paraffin and borated polyethylene (BPE) have been explored for neutron attenuation. The primary reason for selecting these materials is that they are well studied and highly effective for neutron shielding. For the radiogenic neutron spectra present at JUSL \cite{GHOSH2022102700}, attenuation through 10 cm thickness of various materials is studied. The neutron energy spectrum extends from 0 to 10 MeV, with approximately 92\% of the total flux concentrated below 1 MeV. Within this energy range, HDPE (98.8\%) shows higher neutron attenuation compared to water (83.1\%), paraffin (89\%), PE (90.2\%), and BPE (97.2\%). The transmitted neutron spectra for these materials are presented in Fig. \ref{fig4}. Since the attenuation is maximum for HDPE, HDPE is chosen for the neutron shielding design for InDEx.

\section{Simulation work for InDEx}\label{sec3}
InDEx detectors are superheated liquid detectors where an active liquid, C\(_2\)H\(_2\)F\(_4\), is suspended as tiny droplets within a viscoelastic gel matrix. As the gel matrix contains approximately 80\% glycerol, during the simulation, it is considered that the detectors are to be filled with glycerol only. A plane is considered at the centre of the detector, and using USRBDX scoring, the surface crossing neutron fluence is measured. Here, surface crossing neutron fluence means the neutrons that pass perpendicularly through a unit surface area considered at the centre of the detector. Two experiments are simulated, one for a calibration experiment with a \(^{241}\)AmBe (10 mCi) neutron source and the other for run 2 of InDEx with the JUSL neutron background \cite{GHOSH2022102700}. The count rate in the detector is calculated with the surface crossing neutron flux inside the detector using Eq. (\ref{eq1}) \cite{SAHOO2021165450}. 

\begin{equation}
R(E_n,T) = {\phi(E_n)V_l{\sum^{N}_{i}}\epsilon^i(E_n,T)N^i\sigma_n^i(E_n) }
\label{eq1}
\end{equation}
where,
\(\phi(E_n)\) = neutron flux at an energy \(E_n\) incident on the detector active volume, \(V_l\) = detector active liquid volume, \(N^i\) = atomic number density of \(i^{th}\) element of active liquid, \(\sigma_n^i(E_n )\) = neutron interaction cross section of \(i^{th}\) element of active liquid at energy \(E_n\), \(\epsilon^i\) = detector efficiency for \(i^{th}\) element of active liquid \cite{PhysRevD.101.103005}. 

During the rate calculation, neutron fluxes above 7 keV and 11 keV are considered since these energies are the threshold neutron energy of detection at the 1.95 keV detector threshold, as mentioned earlier. 

\begin{figure}
\centering
\includegraphics[width=0.9\textwidth]{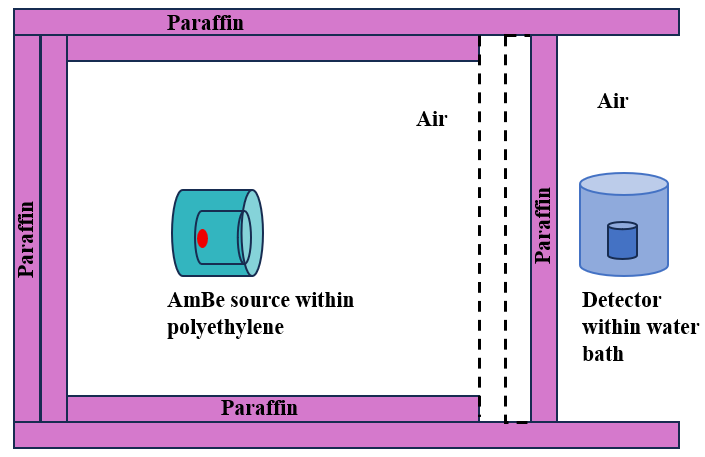}
\caption{The geometry considered in simulation for experimental setup with paraffin blocks.}\label{fig5}
\end{figure}

\begin{figure}
\centering
\includegraphics[width=0.9\textwidth]{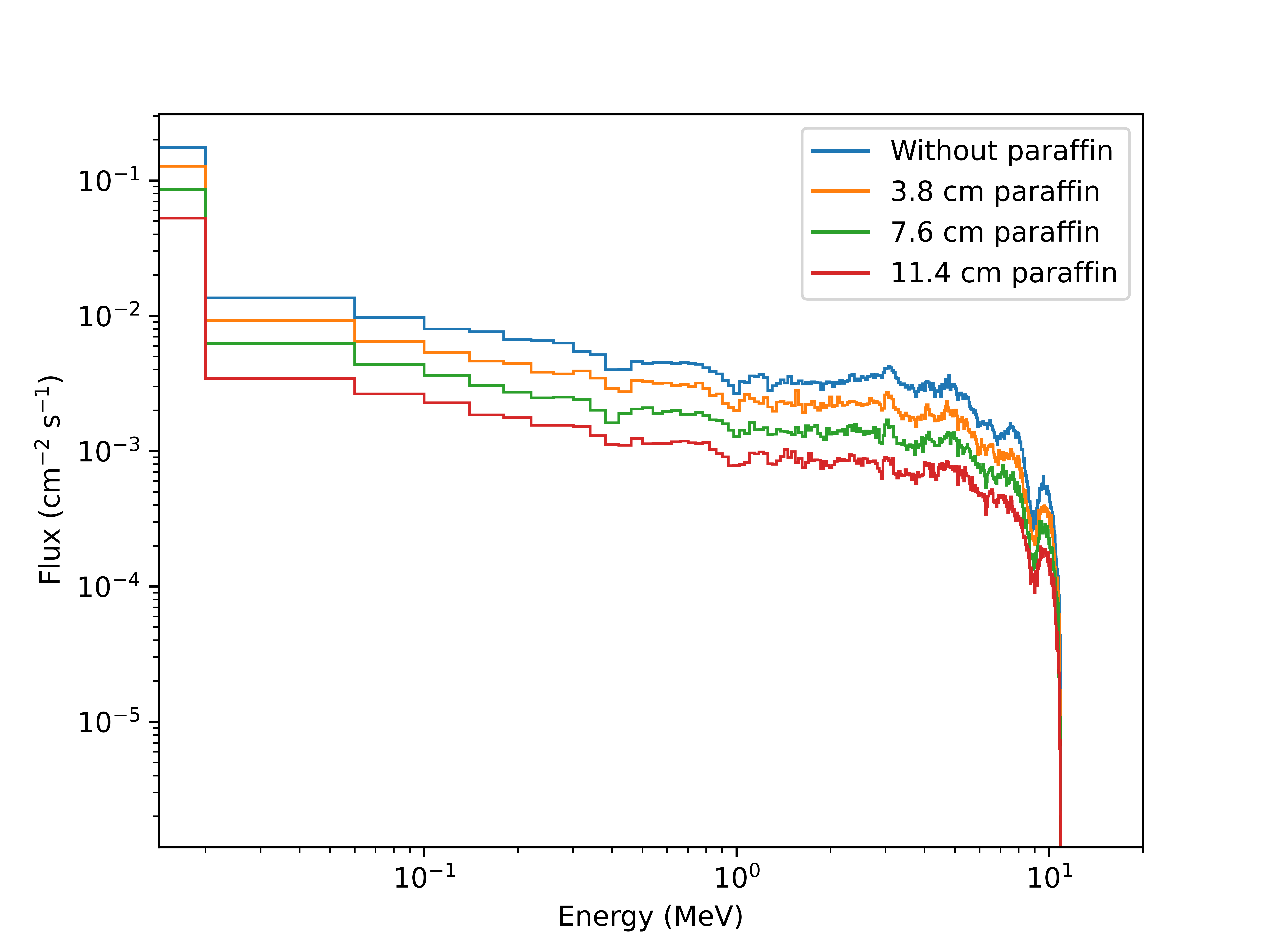}
\caption{FLUKA generated neutron flux distribution inside the detector for different detector geometries as mentioned in Fig. \ref{fig5} with increasing paraffin thickness.}\label{fig6}
\end{figure}

The simulation geometry of the calibration experiments of InDEx using a \(^{241}\)AmBe source and with and without paraffin is shown in Fig. \ref{fig5}. In run 1 and run 2 of InDEx, the calibration experiments were done without any shielding \cite{vpb7-kdrz, KUMAR2026171101}. To observe the effect of shielding in future runs, identical paraffin blocks are used in the calibration experiment. The dimension of the blocks does not represent any specific detector geometry in this case. The distance between the \(^{241}\)AmBe source and detector is 54 cm. The detector is within a cylindrical water bath of borosilicate glass with dimensions (radius \(\times\) height) 5 cm \(\times\) 13 cm. The detector container (cylindrical borosilicate glass) is filled with glycerol having dimensions (radius \(\times\) height) of 2 cm \(\times\) 4 cm. All borosilicate glass containers have a thickness of 0.5 cm. The \(^{241}\)AmBe source is encapsulated within a 5 cm polyethylene shielding. The paraffin blocks used for shielding the detector as well as the surrounding of the source, have dimensions of 60 cm \(\times\) 30 cm \(\times\) 3.8 cm. In simulations, the attenuation of neutron flux due to different paraffin thicknesses (0 -11.4 cm) is explored, and the simulated flux is shown in Fig. \ref{fig6}. The detector count rate is then calculated for the transmitted (attenuated) flux. The attenuated neutron flux and the estimated count rate, along with the experimental event rate, are shown in Table \ref{tab1}. The experimental results and estimated rates are in good agreement for small paraffin thickness and differ when the paraffin thickness is increased. 

\begin{table}
\caption{The attenuated neutron flux and the estimated detector event rate with paraffin shielding.}\label{tab1}%
\begin{tabular*}{\textwidth}{@{\extracolsep\fill}lccccc}
\toprule
 Paraffin   & Simulated total  & Simulated  & Estimated  & Experimental\\
thickness & flux  & flux $\geq$ 7 keV & rate &  rate \\
(cm) & (cm\(^{-2}\) s\(^{-1}\)) & (cm\(^{-2}\) s\(^{-1}\)) & (s\(^{-1}\) g\(^{-1}\)) &  (s\(^{-1}\) g\(^{-1}\))\\
\midrule
 0   & \((7.94\pm0.01)\times10^{-1}\) & \(0.64\pm0.03\) & \((3.63\pm0.28)\times10^{-3}\) & \((3.73\pm0.23)\times10^{-3}\)  \\
 3.8  & \((5.37\pm0.01)\times 10^{-1}\)  & \(0.43\pm0.03\)  & \((2.39\pm0.24)\times10^{-3}\) & \((2.01\pm0.19)\times10^{-3}\)  \\
7.6   & \((3.50\pm0.01)\times10^{-1}\)  & \(0.28\pm0.02\) & \((1.55\pm0.21)\times10^{-3}\) & \((1.15\pm0.18)\times10^{-3}\)  \\
11.4   & \((2.14\pm0.01)\times10^{-1}\)  & \(0.17\pm0.01\) & \((9.50\pm0.10)\times10^{-4}\) & \((4.13\pm0.74)\times10^{-4}\)  \\
\botrule
\end{tabular*}
\end{table}

\begin{figure}
\centering
\includegraphics[width=0.9\textwidth]{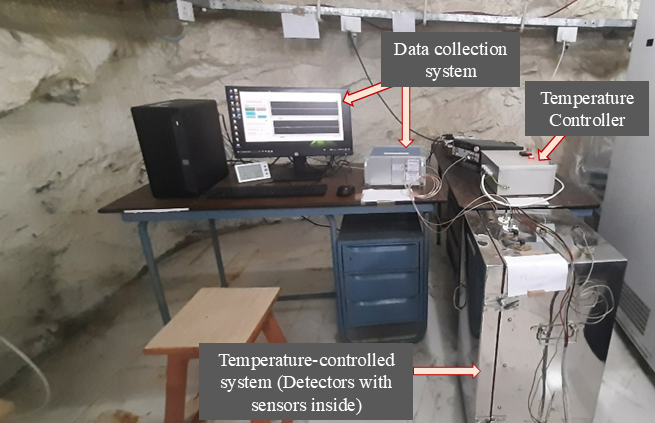}
\caption{The experimental setup of run 2 of InDEx at JUSL.}\label{fig7}
\end{figure}

\begin{figure}
\centering
\includegraphics[height=7.5cm, width=0.5\textwidth]{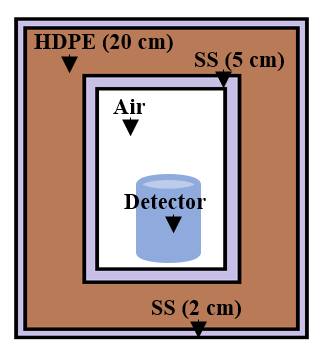}
\caption{Proposed neutron shielding design for InDEx run 2 at JUSL.}\label{fig8}
\end{figure}

The experimental setup of InDEx (InDEx run 2) is simulated next with FLUKA for the neutron background present at JUSL. In this setup, the detector is placed within a temperature-controlled system. The experimental setup of run 2 of InDEx at JUSL is shown in Fig. \ref{fig7}. The temperature controlled system is a stainless steel (SS) box with inner dimensions of 30 cm \(\times\) 30 cm \(\times\) 60 cm and a thickness of 5 cm. The borosilicate glass container of the detector is of dimensions (radius \(\times\) height) of 4 cm \(\times\) 11 cm with a glass thickness of 0.5 cm. During the simulation, the detector is considered to be filled with glycerol only, as mentioned earlier. The neutron source is the background neutron spectrum (radiogenic and cosmogenic) at JUSL, which is extracted from Ref. \cite{GHOSH2022102700}. A parallel neutron beam is projected on the SS surface and the attenuated neutron fluence within the detector is simulated. The detector event rate has been estimated for both radiogenic and cosmogenic neutrons with this attenuated flux using Eq. (1). The rate for radiogenic neutrons comes as \((1.91\pm0.10) \times 10^{-7}\) s\(^{-1}\) g\(^{-1}\) and for cosmogenic neutrons it is \((2.38\pm0.09) \times 10^{-11}\) s\(^{-1}\) g\(^{-1}\). The experimental rate for run 2 of InDEx at JUSL is \((1.37\pm0.15) \times 10^{-7}\) s\(^{-1}\) g\(^{-1}\). 

For future experiments of InDEx, the background neutron suppression or elimination is essential. For this purpose, a shielding of 20 cm thick HDPE is proposed to be placed around the temperature-controlled system. The shielding thickness is optimised to an extent such that the resulting event rate due to background neutrons does not interfere with the experimental measurements. HDPE is placed inside a 2 cm thick stainless steel box. The suggested arrangement for this shielding is shown in Fig. \ref{fig8}. It is observed that the detector event rate due to shielding is reduced to \((2.6\pm0.43) \times 10^{-10}\) s\(^{-1}\) g\(^{-1}\). 

\section{Results and discussion}\label{sec4}

In this work, the neutron shielding for the future InDEx experiment is proposed using the FLUKA simulation. For the validation of results from FLUKA, both monoenergetic and polyenergetic neutron sources alongside polyethylene shielding configurations are evaluated. The calibration experiment with the \(^{241}\)AmBe neutron source and paraffin shielding is simulated and the detector event rate for that experiment is calculated. The calculated event rates agree with experimental results within 20\% for paraffin thickness up to 3.8 cm, but the calculated values are higher than the measured ones when shielding is thicker. One of the possible reasons for this is the detection threshold of the data acquisition (DAQ) system (20 mV), which is not considered during the estimation of event rate. As the thickness of paraffin increases, the higher energy neutron is shifted to lower energy and the DAQ detection threshold may have overlooked these low energy neutron-induced events. In another work, the simulation is carried out with the geometry of InDEx run 2 for the background neutron flux at JUSL. In this case, the estimated count rate is about 39\% higher than the experimental rate. This may depend on a similar reason for the data acquisition system threshold. The primary goal is to minimise the background neutron-induced events in the detector for future InDEx runs. The simulation with HDPE of 20 cm thickness is simulated for this purpose and found suitable as shielding for neutrons. The results show that HDPE can reduce the background event rate of the detector by almost three orders of magnitude. This reduction will help in increasing the detector sensitivity in future runs of InDEx. In the future experiment at 35\(^{\circ}\)C, the exposure is expected to increase to 100 kg-days. The calculated neutron count rate with the simulated neutron flux after 20 cm HDPE shielding is \((2.6\pm0.43) \times 10^{-10}\) s\(^{-1}\) g\(^{-1}\). Therefore, for the next run at that threshold, the expected count for neutron background will be $2.25\pm0.37$ events.

\section{Conclusion}\label{sec5}
The current simulation study was motivated primarily by a need to reduce background in the dark matter experiment, InDEx. FLUKA is a well-known shielding simulation toolkit, and it has been used for neutron shielding design in the InDEx experiment. For previous InDEx experimental runs, the simulated (FLUKA) and experimental results agree satisfactorily. It indicates that the present simulation work for neutron shielding design using HDPE with a thickness of 20 cm will be equally effective in subsequent runs. In the next phase, InDEx will run with HDPE shielding with a reduced neutron background, which will provide a much better sensitivity of the dark matter search for the InDEx experiment.
 
\bmhead{Acknowledgements}
The authors are thankful to Mr Nilanjan Biswas and Ms Soma Roy, SINP, for their assistance during the experiment. The authors are also thankful to HPU, VECC for providing the \(^{241}\)AmBe source. The work is supported by the DAE, GoI, SINP/4001 project fund.

\bibliography{sn-bibliography}

@article{ROY2001271,
title = {{Superheated liquid and its place in radiation physics}},
journal = {Radiation Physics and Chemistry},
volume = {61},
number = {3},
pages = {271-281},
year = {2001},
note = {8th International Symposium on Radiation Physics - ISRP8},
issn = {0969-806X},
doi = {https://doi.org/10.1016/S0969-806X(01)00250-X},
url = {https://www.sciencedirect.com/science/article/pii/S0969806X0100250X},
author = {S.C Roy}
}

@article{PhysRevLett.108.201302,
  title = {{Final Analysis and Results of the Phase II SIMPLE Dark Matter Search}},
  author = {Felizardo, M. and others},
  collaboration = {The SIMPLE Collaboration},
  journal = {Phys. Rev. Lett.},
  volume = {108},
  issue = {20},
  pages = {201302},
  numpages = {5},
  year = {2012},
  month = {May},
  publisher = {American Physical Society},
  doi = {10.1103/PhysRevLett.108.201302},
  url = {https://link.aps.org/doi/10.1103/PhysRevLett.108.201302}
}

@misc{fustin2024darkmatterlimitscoupp,
      title={{First Dark Matter Limits from the COUPP 4kg Bubble Chamber at a Deep Underground Site}}, 
      author={Drew Fustin},
      year={2024},
      eprint={2401.07384},
      archivePrefix={arXiv},
      primaryClass={astro-ph.CO},
      url={https://arxiv.org/abs/2401.07384}, 
}

@Article{Antonicci2017,
author={Antonicci, A.
and others},
Collaboration={{The MOSCAB Collaboration}},
title={{MOSCAB: a geyser-concept bubble chamber to be used in a dark matter search}},
journal={Eur. Phys. J. C},
year={2017},
month={Nov},
day={09},
volume={77},
number={11},
pages={752},
issn={1434-6052},
doi={10.1140/epjc/s10052-017-5313-8},
url={https://doi.org/10.1140/epjc/s10052-017-5313-8}
}

@article{BEHNKE201785,
title = {{Final results of the PICASSO dark matter search experiment}},
journal = {Astropart. Phys.},
volume = {90},
pages = {85-92},
year = {2017},
issn = {0927-6505},
doi = {https://doi.org/10.1016/j.astropartphys.2017.02.005},
url = {https://www.sciencedirect.com/science/article/pii/S092765051730066X},
author = {E. Behnke and others}
}

@article{PhysRevD.100.022001,
  title = {{Dark matter search results from the complete exposure of the PICO-60 ${\mathrm{C}}_{3}{\mathrm{F}}_{8}$ bubble chamber}},
  author = {Amole, C. and others},
  collaboration = {PICO Collaboration},
  journal = {Phys. Rev. D},
  volume = {100},
  issue = {2},
  pages = {022001},
  numpages = {9},
  year = {2019},
  month = {Jul},
  publisher = {American Physical Society},
  doi = {10.1103/PhysRevD.100.022001},
  url = {https://link.aps.org/doi/10.1103/PhysRevD.100.022001}
}

@article{vpb7-kdrz,
  title = {{Dark matter direct search result from InDEx run2 at JUSL}},
  author = {Das, Susmita and others},
  journal = {Phys. Rev. D},
  volume = {112},
  issue = {4},
  pages = {042003},
  numpages = {7},
  year = {2025},
  month = {Aug},
  publisher = {American Physical Society},
  doi = {10.1103/vpb7-kdrz},
  url = {https://link.aps.org/doi/10.1103/vpb7-kdrz}
}

@article{SAHOO201944,
title = {{The threshold of gamma-ray induced bubble nucleation in superheated emulsion}},
journal = {Nucl. Instrum. Methods Phys. Res. Sect. A},
volume = {931},
pages = {44-51},
year = {2019},
issn = {0168-9002},
doi = {https://doi.org/10.1016/j.nima.2019.04.010},
url = {https://www.sciencedirect.com/science/article/pii/S0168900219304589},
author = {Sunita Sahoo and others}
}

@article{SAHOO2021165450,
title = {{The background study at 555 m deep underground with superheated emulsion detector}},
journal = {Nucl. Instrum. Methods Phys. Res. Sect. A},
volume = {1008},
pages = {165450},
year = {2021},
issn = {0168-9002},
doi = {https://doi.org/10.1016/j.nima.2021.165450},
url = {https://www.sciencedirect.com/science/article/pii/S0168900221004356},
author = {Sunita Sahoo and others}
}

@article{KUMAR2026171101,
title = {{Response of tetrafluoroethane (C\(_2\)H\(_2\)F\(_4\)) superheated emulsion detector for dark matter search at JUSL}},
journal = {Nucl. Instrum. Methods Phys. Res. Sect. A},
volume = {1083},
pages = {171101},
year = {2026},
issn = {0168-9002},
doi = {https://doi.org/10.1016/j.nima.2025.171101},
url = {https://www.sciencedirect.com/science/article/pii/S0168900225009039},
author = {V. Kumar and others}
}

@article{PhysRevD.101.103005,
  title = {{Probing low-mass WIMP candidates of dark matter with tetrafluoroethane superheated liquid detectors}},
  author = {Seth, Susnata and others},
  journal = {Phys. Rev. D},
  volume = {101},
  issue = {10},
  pages = {103005},
  numpages = {12},
  year = {2020},
  month = {May},
  publisher = {American Physical Society},
  doi = {10.1103/PhysRevD.101.103005},
  url = {https://link.aps.org/doi/10.1103/PhysRevD.101.103005}
}

@article{GHOSH2022102700,
title = {{Measurements of gamma ray, cosmic muon and residual neutron background fluxes for rare event search experiments at an underground laboratory}},
journal = {Astropart. Phys.},
volume = {139},
pages = {102700},
year = {2022},
issn = {0927-6505},
doi = {https://doi.org/10.1016/j.astropartphys.2022.102700},
url = {https://www.sciencedirect.com/science/article/pii/S0927650522000147},
author = {Sayan Ghosh and others}
}

@article{ALMISNED2024111585,
title = {{Neutron transmission analysis in borated polyethylene, boron carbide, and polyethylene: Insights from MCNP6 simulations}},
journal = {Radiat. Phys. Chem.},
volume = {218},
pages = {111585},
year = {2024},
issn = {0969-806X},
doi = {https://doi.org/10.1016/j.radphyschem.2024.111585},
url = {https://www.sciencedirect.com/science/article/pii/S0969806X2400077X},
author = {Ghada Almisned and others}
}

@article{10.1063/10.0002041,
    author = {Barbagallo, Dylan and others},
    title = {{Neutron Attenuation in Polyethylene Using an AmBe Source}},
    journal = {J. Undergrad. Rep. Phys.},
    volume = {30},
    number = {1},
    pages = {100001},
    year = {2020},
    month = {01},
    issn = {2642-7451},
    doi = {10.1063/10.0002041},
}

@article{article,
author = {Das, Mala and others},
year = {2010},
month = {10},
pages = {675-682},
title = {{Nucleation efficiency of R134a as a sensitive liquid for Superheated drop emulsion detector}},
volume = {75},
journal = {Pramana},
doi = {10.1007/s12043-010-0147-z}
}

@article{PhysRevD.73.053004,
  title = {{Muon-induced background study for underground laboratories}},
  author = {Mei, D.-M. and others},
  journal = {Phys. Rev. D},
  volume = {73},
  issue = {5},
  pages = {053004},
  numpages = {18},
  year = {2006},
  month = {Mar},
  publisher = {American Physical Society},
  doi = {10.1103/PhysRevD.73.053004},
  url = {https://link.aps.org/doi/10.1103/PhysRevD.73.053004}
}

@article{PhysRevD.110.063006,
  title = {{Cosmic ray muons in laboratories deep underground}},
  author = {Woodley, William and others},
  journal = {Phys. Rev. D},
  volume = {110},
  issue = {6},
  pages = {063006},
  numpages = {18},
  year = {2024},
  month = {Sep},
  publisher = {American Physical Society},
  doi = {10.1103/PhysRevD.110.063006},
  url = {https://link.aps.org/doi/10.1103/PhysRevD.110.063006}
}

@article{SHARAN2021165083,
title = {{Measurement of cosmic-ray muon flux in the underground laboratory at UCIL, India, using plastic scintillators and SiPM}},
journal = {Nucl. Instrum. Methods Phys. Res. Sect. A},
volume = {994},
pages = {165083},
year = {2021},
issn = {0168-9002},
doi = {https://doi.org/10.1016/j.nima.2021.165083},
url = {https://www.sciencedirect.com/science/article/pii/S016890022100067X},
author = {Manoj K. Sharan and others}
}

@article{ refId0,
	author = {Ballarini, Francesca and others},
	title = {{The FLUKA code: Overview and new developments}},
	DOI= "10.1051/epjn/2024015",
	url= "https://doi.org/10.1051/epjn/2024015",
	journal = {EPJ Nuclear Sci. Technol.},
	year = 2024,
	volume = 10,
	pages = "16",
}

@article{BOHLEN2014211,
title = {{The FLUKA Code: Developments and Challenges for High Energy and Medical Applications}},
journal = {Nuclear Data Sheets},
volume = {120},
pages = {211-214},
year = {2014},
issn = {0090-3752},
doi = {https://doi.org/10.1016/j.nds.2014.07.049},
url = {https://www.sciencedirect.com/science/article/pii/S0090375214005018},
author = {T.T. Böhlen and others}
}

@article{ refId02,
	author = {Donadon, André and others},
	title = {{FLAIR3 – recasting simulation experiences with the Advanced Interface for FLUKA and other Monte Carlo codes}},
	DOI= "10.1051/epjconf/202430211005",
	url= "https://doi.org/10.1051/epjconf/202430211005",
	journal = {EPJ Web Conf.},
	year = 2024,
	volume = 302,
	pages = "11005",
}

\end{document}